\documentstyle[prd,aps,preprint]{revtex}
\begin{document}
\draft

\title{The Phases and Triviality of Scalar Quantum Electrodynamics}

\author{M.~Baig and H.~Fort}
\address{Grup de F\'{i}sica Te\`orica,
Institut de F\'{i}sica d'Altes Energies, Universitat Aut\`onoma de
Barcelona, 08193 Bellaterra (Barcelona) SPAIN}
\author{J.~B.~Kogut}
\address{Physics Department, 1110 West Green Street,
University of Illinois,
Urbana, IL  61801-3080}
\author{S.~Kim}
\address{ High Energy
Physics Division, Argonne National Laboratory, Argonne, IL  60439}

\date{April, 1994}

\maketitle

\begin{abstract}

The phase diagram and critical behavior of scalar quantum
electrodynamics  are investigated using lattice gauge theory
techniques. The lattice action fixes the length of
the scalar (``Higgs'') field and treats the gauge field as non-compact.
The phase diagram is two dimensional. No fine tuning or extrapolations
are needed to study the theory's critical behovior. Two lines of second
order phase transitions are discovered and the scaling laws for each
are studied by finite size scaling methods on lattices ranging from
$6^4$ through $24^4$. One line  corresponds to monopole percolation
and the other to a transition between a ``Higgs'' and a ``Coulomb''
phase, labelled by divergent specific heats. The lines of transitions
cross in the interior of the phase diagram and appear to be unrelated.
The monopole percolation transition has critical indices which are
compatible with ordinary four dimensional percolation uneffected by
interactions. Finite size scaling and histogram methods reveal that the
specific heats on the ``Higgs-Coulomb'' transition line are well-fit by
the hypothesis that scalar quantum electrodynamics is logarithmically
trivial. The logarithms are measured in both finite size scaling of
the specific heat peaks as a function of volume as well as in the
coupling constant dependence of the specific heats measured  on fixed
but large lattices. The theory is seen to be qualitatively similar to
$\lambda\phi^{4}$.

The standard CRAY random number generator RANF proved to be inadequate
for the $16^4$ lattice simulation. This failure and our
``work-around'' solution are briefly discussed.

\end{abstract}


\newpage
\section{Introduction}

{}~~~~~~In a recent letter$^{1.}$ we presented a lattice gauge theory study
of scalar quantum electrodynamics (SQED) which provided strong numerical
evidence for the logarithmic triviality of the theory. It is the
purpose of this paper to both provide further detail underlying that
letter, as well as present a more comprehensive view of SQED by
discussing additional lattice calculations. These new calculations
will include monopole percolation observables, the coupling constant
dependence of the model's specific heat, evidence for logarithms of
triviality in the finite size scaling variable of the model's
specific heat peaks
and a simulation of the four
dimensional planar spin model.  We shall see that there is a
line of monopole percolation transitions
in the phase diagram of SQED, but {\it unlike
fermionic} lattice QED, it does not coincide with the bulk transition
separating the Higg's and Coulomb phases of the model and is,
therefore, irrelevant to the theory's continuum limit. We will
investigate the theory's continuum limit for a fairly large value of
the bare gauge coupling. As already reported in ref.1, the
Higg's-Coulomb phase transition  will prove to be compatible with a
logarithmically trivial continuum theory. Finite size scaling studies
of the specific heat peaks and their positions in the phase diagram as
a function of lattice volume, point to  logarithmically improved mean
field theory as an accurate effective field theory. The correlation
length exponent $\nu$ is 0.50(2), which is compatible with the free
field result of $1/2$. The specific heat peaks do grow with lattice
size, but the data strongly favor a slow logarithmic volume dependence
rather that the power law dependence expected of a non-trivial
continuum theory. New measurements of the dependence of the specific
heats on the bare coupling constants also expose logarithmic
modifications of pure mean field predictions. In fact, this study
supports the view that SQED has scaling behavior which is qualitatively
similar to $\lambda\phi^{4}$.

There are several theoretical as well as phenomenological motivations
for this work. On the theory side, the search continues for an
interacting ultra-violet fixed point field theory in four dimensions.
Our numerical evidence suggests that SQED suffers from the zero charge
problem$^{2.}$ like $\lambda\phi^{4}$. Another theoretical motivation for
this work is our recent investigation of {\it fermionic} QED whose
simulation results could be fit with the scaling laws of a non-trivial
field theory with an ultra-violet stable fixed point$^{3.}$. It was
also observed in those simulations that monopole percolation is
coincident with the chiral transition and that the chiral transition
has the same correlation length index $\nu$ as four dimensional
percolation. These two coincidences have led to the speculation that
monopole percolation is ``driving'' a non-trivial chiral transition in
lattice fermion QED and this is leading to an interacting
ultra-violet fixed point. The spin $1/2$ character of the fermion is
essential in this physical picture because a percolating network of
monopoles can induce rapid helicity flips leading to chiral symmetry
breaking and the index $\nu$ of the monopole network could be
inherited by the chiral transition$^{3.}$. We do not expect such sensitivity
to monopoles in SQED, and we shall find that the Higgs-Coulomb
transition appears to be unrelated to monopole percolation since the
transition lines for each phenomena are separate and actually cross in
the interior of the model's two dimensional phase diagram.

This article is organized into several sections. In Sec.~2 we map out
the two dimensional phase diagram of the model, and show that the
Higgs-Coulomb line is separate from the monopole percolation line. The
main concepts and observables of monopole percolation are briefly
reviewed.  In Sec.~3 we present the finite size scaling data and
analysis of the specific heat peak characterizing the Higgs-Coulomb
transition at a fixed, large gauge coupling. The logarithmic growth of
the peak is quite clear in the data. In addition, the finite size
dependence of the critical coupling is well fit with a correlation
length index $\nu=.50(2)$ which is compatible with a free field
description of the transition.  A careful study of the Binder Cumulant
and related moments of the specific heat data confirms that the
transition is second order. No evidence is found for a weak or
fluctuation-induced transition that bedevil other lattice studies of
the Higgs model. This result is a clear advantage of the non-compact
gauge/fixed length Higgs field formulation used here. Finally, we
present new specific heat measurements on $12^4$, $16^4$, and $20^4$
lattices which show further evidence for logarithmic triviality.
In particular, using histogram methods, we obtained the shape of the
specific heat peaks on each lattice. The resulting curves could be
mapped onto a universal specific heat curve if scale breaking logarithms
are incorporated into the otherwise gaussian model finite size scaling
variable. This analysis is inspired by recent work in $\lambda\phi^{4}$
models and is, to our knowledge, the first quantitatively interesting
study of its kind.
In Sec.~4 we present the data and analysis
of a nearby point on the monopole percolation line of transitions. The
data are consistent with the scaling laws of four dimensional
percolation. In fact, we find that the ratio of the monopole
susceptibility index $\gamma$ to the monopole correlation length index
$\nu$ is 2.25(1), in perfect agreement with the presumably exact
result $9/4$. In Sec.~5 we present data and analysis of the limit of
SQED where the gauge coupling is set to zero and the model reduces to
the $O(2)$ spin model. In this limit mean field theory modified
by calculable scale breaking logarithms should apply. The point of this
exercise is not to present yet another study of the fixed length
$O(2)$ model, but simply to check that the methods, lattice sizes, and
statistics used in the rest of this work can reproduce known answers.
Finally, in Sec.~6 we present some conclusions and suggestions for
related work.

\section{Phase Diagram and Overview}

{}~~~~~~We begin with a lattice formulation of scalar electrodynamics which
is particularly well suited for numerical work and can make contact with
continuum physics with a minimum of fine tuning.  Consider the
non-compact formulation of the abelian Higgs model with a fixed length
scalar field,$^{4.}$

$$
S=\frac{1}{2}\beta\sum_{p}\theta_{p}^{2}-\gamma\sum_{x,\mu}(\phi_{x}^{\ast}
U_{x,\mu} \phi_{x+\mu}+c.c.)
\eqno(1) $$

\noindent where p denotes plaquettes, $\theta_{p}$ is the circulation
of the non-compact gauge field $\theta_{x,\mu}$ around a plaquette,
$\beta=1/e^{2}$ and $\phi_{x}=exp(i\alpha(x))$ is a phase factor at
each site.  We choose this action (the electrodynamics of the planar
model) because preliminary work has suggested that it has a line of
second order transitions,$^{4.}$ because it does not require fine
tuning and because it is believed to lie in the same universality
class as the ordinary lattice abelian Higgs model with a conventional,
variable length scalar field.$^{5.}$  In Fig.~1 we show the phase
diagram of the model in the bare parameter space $\beta-\gamma$.  In
the ``Higgs'' region of the phase diagram the gauge field develops a
mass dynamically, while in the ``Coulomb'' phase it does not.  Earlier
work on this model indicates that the phase transition shows up
clearly in the model's internal energies.  A preliminary investigation
has indicated that the line emanating from the
$\beta\rightarrow\infty$ limit of Fig.~1 is a line of critical points
which potentially could produce a family of interacting, continuum
field theories.$^{4.}$  Note that in the $\beta\rightarrow\infty$
limit the gauge field in Eq.~(1) reduces to a pure gauge
transformation so the model becomes the four dimensional planar model
which is known to have a second order phase transition which is
trivial, i.e. is described by a free field. The $\gamma\rightarrow\infty$
limit of the transition line is also interesting and was discussed
briefly in Ref. 4.
The non-compact nature of
the gauge field is important in Fig.~1---the compact model has a line
of first order transitions and only at the endpoint of such a line in
the interior of a phase diagram can one hope to have a critical point
where a continuum field theory might exist.$^{6.}$  Since one must
fine tune bare parameters to find such a point, the compact
formulation of the model is much harder to use for quantitative
work.$^{6.}$  The fact that Eq.~(1) uses fixed length scalar fields
avoids another fine tuning---the variable length scalar field
formulation would possess a quadratically divergent bare mass
parameter which would have to be tuned to zero with extraordinary
accuracy to search for critical behavior.  Conventional wisdom based
on the renormalization group states that Eq.~(1) should have the same
critical behavior as the fine-tuned, variable length model,$^{5.}$ so
it again emerges as preferable.  Note also that in the naive classical
limit where the field varies smoothly, Eq. (1) reduces to a free
massive vector boson.  In the vicinity of the strong coupling critical
point we investigate here, the fields are rapidly varying on the scale
of the lattice spacing and we shall see that the specific heat scaling
law is {\it not} that of a Gaussian model.

In order to map out the phase diagram we first measured the internal
energies,

$$
E_{\gamma}=\frac{1}{2}<\sum_{p} \theta_{p} ^{2}>
,\hspace{.5in} E_{h} = <\sum_{x,\mu} \phi_{x} ^{\ast} U_{x,\mu}
\phi_{x+\mu} + c.c.> \eqno(2) $$

\noindent on small lattices. This approach has been used in the past
to map out the Higgs-Coulomb phase transition line in similar
models$^{5.}$.

To locate the line of Higgs-Coulomb transitions we made ``heating''
and ``cooling'' runs along lines of fixed $\beta$ or $\gamma$ on a
$6^4$ lattice.  Typically, 250 iterations were done at each coupling
and measurements were taken.  Then the relevant coupling was changed
by $\pm$ 0.005, and another 250 iterations were made, etc.  These
hystersis runs were repeated with greater statistics in several
cases.  For example, the results at $\beta =$0.1 and 0.2 shown in
Fig.~2 and 3 resulted from runs with 6,000 iterations per point.
Both internal energies suggest a Higgs-Coulomb phase transition at
$\gamma=.35$ when $\beta=0.1$, and both suggest that it moves to
$\gamma=.25$ when $\beta$ is set to $0.2$. Runs of this sort were also
done over a wide range of $\gamma$ and $\beta$ values, and the two
dimensional contour plots of the internal energies shown in Fig.~4 and
5 resulted. The resulting line of the Higgs-Coulomb transitions is
shown in Fig.~6 as the continuous, dark line. This crude map of the
phase diagram will prove very helpful in guiding the large scale
simulations which will be described below. We will confirm that the
Higgs-Coulomb phase transition is second order and is compatible with
the logarithmic triviality of SQED.

It would be interesting to understand the dynamics behind the
Higgs-Coulomb phase transition in this model.  The limiting case of
the model when $\beta$ approaches infinity (the gauge coupling $g^2$
vanishing) corresponds to the four dimensional planar spin model which
experiences an order-disorder transition in $\gamma$ which is described
by mean field theory modified by calculable scale breaking logarithms.
The major issue in this investigation is whether
this transition becomes nontrivial as the gauge coupling is taken
different from zero and we move inside the phase diagram of Fig.~6.
Long range vector forces are certainly capable of doing this and there
are many examples of similar phenomena in the statistical mechanics
literature. Four dimensional model field theories, such as the
{\it gauged} Nambu-Jona Lasinio model solved in the ladder
approximation, have analogous behavior$^{7.}$: when the {\it gauge}
coupling is set to zero the model reduces to the pure  Nambu-Jona
Lasinio model which has a (chiral) transition which is trivial, but
when the gauge coupling is nonzero the theory develops anomalous
dimensions which grow with $g^2$. Of course, the gauged Nambu-Jona
Lasinio model has not been solved beyond the ladder approximation so
it is not known if the screening produced by internal fermion loops
reduces the effective gauge coupling to zero rendering the theory
noninteracting. This problem is under active research by lattice
methods using the noncompact formulation of fermionic QED.

In addition to studying the usual order parameters and bulk
thermodynamic quantities in order to search for phase transitions and
scaling laws, it has proved stimulating to also consider the effective
monopole operators introduced by Hands and Wensley$^{8.}$. The reader
should consult the references for background on this extensive
subject, so we will just review some of the essentials here. Even
though the lattice action is noncompact, one can have finite action
monopole loops on the lattice by virtue of the lattice cutoff. These
monopoles are not necessarily physically significant because the pure
gauge action is noncompact and purely gaussian. For example, effective
monopoles can be found  in the quenched model$^{8.}$ which is a free
field and the effective monopoles cannot interact or experience real
dynamics like the monopoles of pure compact lattice gauge theory.
However, as emphasized in ref.8, since matter fields couple to gauge
fields through phase factors which implement the $U(1)$ gauge group,
they could be significant and physical in the full theory. In fact, in
noncompact fermion lattice QED with two or four species, the chiral
transitions are coincident with the monopole percolation transition
and they share the same correlation length scaling index $\nu$$^{3.}$.
These points have led to the inevitable speculation that monopole
percolation is an essential ingredient in the chiral transitions in
the fermion models. These are subjects of active research and many
pieces are missing in the puzzles associated with these ideas.
Nonetheless, it is interesting to look for monopole percolation in SQED
and see if it is related to the Higgs-Coulomb transition found in the
bulk thermodynamics. In fact, we shall find that the two transitions
are {\it not} coincident in the two dimensional phase diagram of SQED.

Monopole percolation is detected using an order parameter and a
susceptibility borrowed from standard percolation models. In this
construction a conserved magnetic current is defined on the dual
lattice exactly as it is done in compact lattice QED$^{9.}$. Then the
idea of a  connected cluster of monopoles is introduced: one counts
the number of dual sites joined into clusters by monopole line
elements. An order parameter for a percolation transition is then
$M=n_{max}/n_{tot}$, where $n_{max}$ is the number of such sites in
the largest monopole cluster and $n_{tot}$ is the total number of
connected sites.  Its associated susceptibility reads,

$$
\chi=\langle\left( \sum_{n} g_{n} n^{2} - n^{2}_{max}
\right) / n_{tot}\rangle \eqno(3) $$

\noindent where $n$ labels the number of sites in a monopole cluster
which occurs $g_{n}$ times on the dual lattice.

The percolation order parameter $M$ and its susceptibility $\chi$ were
then calculated at fixed values of $\gamma$ and variable $\beta$ on a
$10^4$ lattice to search for line(s) of percolation transitions. Our
past experience with quenched noncompact lattice QED as well as the
two and four species models suggested that the line of percolation
transitions would occur near $\beta=0.2$ and be relatively insensitive
to $\gamma$. The simulation gave results in good agreement with these
expectations. The percolation transition line (dashed, with squares)
is shown in Fig.~6.  The squares in Fig.~6 denote the maxima found in
the percolation susceptibility in simulation runs in which $\gamma$
was held fixed on a $10^4$ lattice, and ``heating'' and ``cooling''
runs were made across the peak.  Typically, 4,000 iterations were made
for thermalization at each $\beta$, then an additional 16,000
iterations were made for measurements.  Next, $\beta$ was incremented
by $\pm$ 0.002 and the process was repeated.  Accurate measurements
and finite size scaling studies of $M$ and $\chi$ on larger lattices
will be discussed below.  The line was located from peaks in the
percolation susceptibility and some examples of such  measurements
will be plotted in Sec.~4 where a quantitative finite size scaling
study of the percolation transition will be reported.

The first thing we notice from these measurements is that the
Higgs-Coulomb and the monopole percolation transitions are clearly
distinct and, therefore, unrelated. This result will be confirmed on
much larger lattices. This result stands in sharp contrast to fermion
noncompact lattice QED and suggests that the physics of the phase
transitions in the two models are quite different.

\section{Finite Size Scaling and the Specific Heats}

{}~~~~~~In order to understand the nature of the Higgs-Coulomb phase
transition, we measured critical indices by doing a careful finite
size scaling study of the specific heats related to the internal
energies introduced above. We considered the specific heats
$C_{\gamma} = \partial E_{\gamma}/\partial  \beta$, and $C_{h} =
\partial E_{h}/\partial\gamma$.   In general, singular behavior in
such specific heats at critical couplings can be used to find,
classify, and measure the critical indices of phase transitions.   On
a $L^{4}$ lattice the size dependence of a generic specific heat at a
second order critical point should scale as,$^{10.}$

$$
C_{max} (L) \sim L^{\alpha/\nu}  \eqno(4) $$

\noindent where $\alpha$ and $\nu$ are the usual specific heat and
correlation length critical indices, respectively.  Here $C_{max}$
denotes the peak of the specific heat.  A measurement of the index
$\nu$ can be made from the size dependence of the position of the
peak.  In a model which depends on just one coupling, call it $g$,
then$^{10.}$

$$
g_{c} (L) - g_{c} \sim L^{-1/\nu}  \eqno(5) $$

\noindent where $g_{c} (L)$ is the coupling where $C_{max} (L)$ occurs
and $g_{c}$ is its $L\rightarrow\infty$ thermodynamic limit.  The
scaling laws Eq. (4) and (5) characterize a critical point with
powerlaw singularities.  This is a possible behavior for scalar
electrodynamics, but there is also the possibility suggested by
perturbation theory, that the theory is logarithmically trivial.
Consider $\lambda \phi^{4}$ as the simplest, well-studied theory which
apparently has this behavior.  In this case the theory becomes trivial
at a logarithmic rate as the theory's momentum space cutoff $\Lambda$
is taken to infinity.  Then the scaling laws of Eq. (4) and (5)
become,$^{11,12}$

$$
C_{max}(L)\sim({\ell}nL)^{p} \eqno(6) $$

\noindent and

$$
g_{c}(L)-g_{c}\sim\frac{1}{L^{2}({\ell}nL)^{q}} \eqno(7) $$

\noindent where $p$ and $q$ are powers predictable in one-loop
perturbation theory ($p=\frac{1}{3}$ and $q=\frac{1}{6}$ in
$\lambda\phi^{4}$).  Note the differences between these scaling laws
and those of the usual Gaussian model, obtained from Eq. (4) and (5)
setting $\alpha=0$ and $\nu=.5$:  in the Gaussian model the specific
heat should saturate as $L$ grows, and the position of the peaks
should approach a limiting value at a rate $L^{-2}$.

It is particularly interesting in scalar electrodynamics to consider a
large value of the bare (lattice) gauge coupling to see if that can
induce non-trivial interactions which survive in the continuum limit.
So, we ran extensive simulations on lattices ranging from $6^{4}$
through $20^{4}$ at $e^{2}=5.0$ and searched in parameter space
$(\beta,\gamma)$ for peaks in $C_{\gamma}$ and $C_{h}$.  We used
histogram methods$^{13,14}$ to do this as efficiently as possible.
For example, on a $6^{4}$ lattice at $\beta=.2000$ and $\gamma=.2350$
we found a specific heat peak near $\gamma_{c}(6)\approx.2382$ from
the histogram method. The peak is shown in Fig.~7.  The $\gamma$
value in the lattice action was then tuned to .2382 and additional
simulations and  histograms produced specific heats, found from the
variances of $E_{\gamma}$ and  $E_{h}$ measurements, at a $\gamma_{c}$
very close to .2382.  Using this strategy, measurements of
$\gamma_{c}(L),C_{\gamma}(L)$ and $C_{h}(L)$ could be made without
relying on any extrapolation methods.  We thus avoided systematic
errors, although  critical slowing down on the larger lattices limited
our statistical accuracy. In Fig.~8 and 9 we show the internal energy
$E_{h}$ and specific heat $C_{h}$ on $12^4$ and $18^4$ lattices,
respectively. Note that the peaks sharpen and shift to smaller
$\gamma$ values as $L$ increases. These effects will be studied
more systematically below when the shapes of each specific heat
curve will be used to detect logarithmic scale breaking.
In Table 1 we show a subset of our
results that will be analyzed and discussed here.  The columns labeled
$\gamma_{c}(L),C_{\gamma}^{max}(L)$ and $C_{h}^{max}(L)$ in Table 1
need no further explanation except to note that the error bars were
obtained with standard binning procedures which account for the
correlations in the data sets produced by Monte Carlo programs.

The Monte Carlo procedure used here was a standard multi-hit
Metropolis for the non-compact gauge degrees of freedom and an
over-relaxed plus Metropolis  algorithm$^{15.}$ for the compact matter
field.  Over-relaxation reduced the correlation times in the algorithm
by typically a factor of 2--3.  Accuracy and good estimates of error
bars are essential in a quantitative study such as this.
Unfortunately, cluster and acceleration algorithms have not been
developed for gauge theories, so very high statistics of our
over-relaxed Metropolis algorithm were essential---tens of millions of
sweeps were accumulated for each lattice size as listed in column 7 of
Table 1. We also wrote a unitary gauge code which eliminated the
matter field entirely from the algorithm. Extensive runs on lattices
ranging from $4^4$ to $16^4$ produced the same observables as the
original code. These results provided an excellent check on the
correctness of our programming and confirmed that our codes properly
converged to statistical equilibrium.

A word of warning for the ambitious---the standard CRAY random number
generator RANF which uses the linear congruent  algorithm with modulus
$2^{48}$ proved inadequate for lattices whose linear dimension was a
power of 2, such as $16^4$.
Presumably this occurred because for strides of length
$2^{N}$ the period of RANF is reduced from $2^{46}$ to $2^{(46-N)}$
and the well-known correlations in such generators are expected to
have maximum effect if the distribution is sampled with a period of
$2^{N}$.  The simplicity of the variables and Monte Carlo algorithm
for lattice scalar electrodynamics also makes
it more susceptible to the correlations in random number generators
than other models.  We discovered this problem when our $16^{4}$
simulations were unstable -- very long runs produced specific heat
peaks that grew without apparent bound and shifted to large $\gamma$.
After considerable investigative work, we
isolated the problem in the random number generator.  We cured the
problem by adding extra calls to RANF to avoid strides of length
$2^{N}$.  Problems with generally accepted random number generators
have been studied systematically in ref.~(16)

Specific heats were measured as the fluctuations in internal energy
measurements $(C_{h}=(<E_{h}^{2}>-<E_{h}>^{2})/4L^{4}$, etc.).  Very
high statistics and many $L$ values are needed to distinguish between
logarithmic triviality (Eq.~6) and powerlaw behavior (Eq.~4).  The
other entries in Table 1, $K_{\gamma}(L)$ and $K_{h}(L)$, are  the
Binder Cumulants (Kurtosis)$^{17.}$ for each internal energy.  At a
continuous phase  transition each Kurtosis should approach $2/3$ with
finite size corrections scaling as  $1/L^{4}$.  The Kurtosis is a
useful probe into the order of a phase transition, although  an
examination of the internal energy and specific heat histograms are
often just as  valuable. Since the order of the transitions in lattice
and continuum scalar  electrodynamics are controversial, we studied
these quantities with some care.

Consider the Kurtosis $K_{\gamma} (L)$, the specific heat
$C_{\gamma}^{max} (L)$ and the critical coupling $\gamma_{c}(L)$ of
scalar electrodynamics.  As stated above, we set the lattice (bare)
gauge coupling to $e^{2} = 5.0$ and then used simulations, enhanced by
histogram methods, to locate the transition line in Fig.~1.  The
Kurtosis $K_{\gamma} (L)$ is plotted against $10^{6}/L^{4}$ in
Fig.~10.  The size of the symbols include the error bars, and clearly
the curve favors a second order transition.  A three parameter fit to
the $L=12$, 14, 16, 18 and 20 data using the form
$K_{\gamma}(L)=aL^{\rho}+b$ is excellent (confidence level = 98\%)
predicting $\rho=-4.1(4)$ and $K_{p}(\infty)=.666665(2)$.  The
hypothesis of a line of second order transitions in Fig.~1 appears to
be very firm, with no evidence for a fluctuation-induced first order
transition.  An analysis of $K_{h} (L)$ gives the same conclusion with
somewhat larger error bars. In Fig.~11 we show $K_{h}(L)$ and find
compatibility with the value $2/3$ for large $L$.

Next we plot our $C_{\gamma}^{max} (L)$ data vs. $L$ in Fig.~12.  We
attempted powerlaw as well as logarithmic finite size scaling
hypotheses.  The powerlaw hypothesis did {\it not} produce a
stable fit for any reasonable range of parameters.  However,
logarithmic fits were quite good.  The hypothesis $C_{\gamma}^{max}
(L) = a {\ell}n^{\rho}L+b$ for $L=8$,10,12,14,16,18 and 20 fit with a
confidence level $=~90\%$ producing the estimate $\rho=1.4(2)$.  If we
considered the range $L=8-18$, the same fitting form predicted $\rho =
1.5 (3)$ with confidence level $=84\%$, and if the range $L=10-20$
were taken we found $\rho=1.4 (5)$ with confidence level $=78\%$.  The
solid line in Fig.~12 is the $L=8-20$ fit.  An analysis of
$C_{h}^{max} (L)$ gave consistent results---the same logarithmic
dependence should be found in either specific heat---and powerlaw fits
to $C_{h}^{max}(L)$ were also ruled out.  In particular, a fit of the
form $C_{h}^{max}(L)=a{\ell}n^{\rho}L+b$  for $L=8-18$ gave $\rho =
0.9 (3)$ with confidence level $=82\%$ and for $L=8-20$ gave
$\rho=1.0(2)$ with confidence level =85\%. We show the data and the
logarithmic fit in Fig.~13.

Next, in Fig.~14 we show $\gamma_{c}(L)$ vs. $10^{4}/L^{2}$.  As $L$
increases the specific heat peak shifts to smaller $\gamma_c(L)$, and
the rate of the shift is determined by the critical index $\nu$ in a
scaling theory.  The error bars again fall within the symbols in the
figure.  The data is clearly compatible with the correlation length index
$\nu = 0.5$ expected of a theory which is free in the continuum limit.
In the case of $\lambda \phi^{4}$ it has proven possible to find the
logarithm of Eq.~(7) under the dominant $L^{-2}$ behavior by using special
techniques.$^{18.}$   We do not quite have the accuracy to do that here:
a powerlaw fit to $\gamma_{c} (L)=\gamma_{c}+a/L^{1/\nu}$ using $L =
12-20$ predicts $1/\nu = 2.0 (1)$, $\gamma_{c}=.22825(8)$ with confidence
level $=92\%$ and using $L=14-20$ predicts $1/\nu=1.9(3)$,
$\gamma_{c}=.2282 (2)$ with confidence level $=97\%$.

Taken together, these measurements of the size dependence of the critical
couplings and specific heat peaks provide good evidence that SQED
is logarithmically trivial. The measurement of $\nu$ is essential
here -- taken on its own, the specific heat data
on the heights of the peaks would be less persuasive
since one can cite models in less than four dimensions with nonzero
anomalous dimensions but with just logarithmically singular or even
nonsingular specific heats. However, in four
dimensions hyperscaling correlates the powers
$\nu = 1/2$ and $\alpha = 0$, and, on the basis of explicit
calculations in $\lambda\phi^{4}$, the modifications of such
scaling laws due to logarithms are rather well understood$^{18.}$.
{}From this point of view, our measurements of $\nu$ and $\alpha$ are
nicely consistent, and suggest that SQED may be logarithmically trivial
in a fashion qualitatively similar to $\lambda\phi^{4}$.

Another insight into the dynamics of the model follows from the
{\it shapes} of the specific heat peaks when plotted against the theory's
bare couplings.  As listed in Table~2 we did additional simulations at
various $\gamma$ values on $12^4, 16^4$ and $20^4$ lattices to obtain
the shapes of the specific heat $C_h(\gamma)$ for $\beta$ fixed at
.2000.  The results are plotted in Fig.~15 where we see that the peaks
move to smaller $\gamma_{c}(L)$ and become narrower as $L$ increases.

In order to get such accurate results we used the multi-histogram
techniques of Ref.~14 to combine data from different $\gamma$
values.  We found that the method was effective only for $\gamma$
values near the $\gamma$ value used in the simulation itself.  In
particular, 1.5 million sweeps of the $16^4$ lattice were made at each
$\gamma$ value listed in Table~2.  The data lists at several $\gamma$
values above and below the one in question ($\gamma_o$, say) were used
to obtain additional predictions for $C_h(\gamma_o)$ and reduce its
uncertainty.  The errors quoted in the table come from standard binning
methods treating each estimate of $C_h(\gamma_o)$ as statistically
independent.  We found that only nearby values of $\gamma$ were useful
in reducing the variances--estimates of $C_h(\gamma_o)$ coming from
simulations at very different $\gamma$ values had too much scatter to
help pinpoint the actual $C_h(\gamma_o)$ value.  Additional statistics
at each $\gamma$ run would certainly improve the utility of the
multi-histogramming methods, as observed by many other authors studying
a wide variety of models.  The histogram method was quite successful
here, nonetheless, and the statistical errors reported in Table~2 are
smaller by a factor of 2-3 as compared to the raw data at each
$\gamma$ value.

The first point we wish to investigate is whether the data of Fig.~15
can be understood from the perspective of finite size scaling.  If
the theory were described by power-law singularities, then the
specific heat data should follow a universal curve,$^{10.}$
$$
C_h(\gamma,L) \sim L^{\alpha /\nu} f (\Delta t) \eqno(8a) $$

\noindent where
$$
\Delta t = (\gamma - \gamma_c(L)) L^{1/\nu} \eqno(8b) $$

\noindent
Clearly Eqs. (8a, b) generalize Eqs. (4) and (5) above.  With scale
breaking logarithms we expect instead$^{18.}$,

$$
C_h(\gamma, L) \sim (\ln L)^{p} f (\Delta t) \eqno(9a) $$

\noindent where
$$
\Delta t = (\gamma - \gamma_c (L) ) L^2 (\ln L)^q
(\ln (\gamma-\gamma_c(L)))^r \eqno(9b) $$

\noindent
In either case, Eq. (8) or Eq. (9), the specific heat peaks should
increase with $L$ and become narrower when plotted against $\gamma$.
These qualitative effects are clear in the data.  Given data on
just three lattice sizes the functional form of the
scaling prefactors in Eqs.~(8a) and (9a)
will not be challenged here, but the widths of the peaks can provide
some insight.  We expect that Eq.~(8b), with $\nu$ set to .50, will be
fairly successful in describing the narrowing of each peak in light of
Fig.~14.  In Fig.~16 we plot the $12^4$, $16^4$ and $20^4$ data in the
form of Eq.~(8) after rescaling the height of each peak to the $12^4$
data, using the more accurate data for the peaks in Table 1.
The ``near universal'' character of the data is clear with $\nu
=$ 1/2 but as $L$ increases the data falls systematically below Eq.~(8).
It is interesting, however, that the scaling form of the data
can be markedly improved by including a logarithm of scale breaking as
suggested by Eq.~(9).  In Fig.~17 we replot the data, scaled to a
common height, using Eq.~(9b) with $q=$ 1 and $r=$ 0.
The curves in Fig.~17 overlap beautifully now, giving good evidence
that logarithmic corrections to gaussian exponents can
accommodate the entire data set.

We can find additional evidence for logarithmic violations of scaling
and triviality by analyzing the $\gamma$ dependence of each peak.  For
infinite $L$ the specific heat should diverge logarithmically in this
scenario,

$$
C_h (\gamma, L \rightarrow \infty) \sim ln^{p^\prime}
\mid\gamma - \gamma_c\mid
\eqno(10) $$

\noindent
On a finite lattice this sort of result is, in general, hard to confirm
because it depends on the existence of a ``scaling window''--for each
$L$ one must find a range of $\gamma$ where Eq.~(10) holds, undistorted
by finite size effects which occur when
$\gamma$ is chosen too close to $\gamma_c$
and undistorted by finite
lattice spacing effects which occur when
$\gamma$ is chosen too far from $\gamma_c$.  We
considered the $16^4$ data and tried fits of the form $C_h(\gamma,
16) = a\, \ln^{p^\prime}\mid\Delta\gamma\mid +$ b
both above and below the peak.
Choosing the points at $\gamma =$ .2300 - .2310 we found a fit with a
65\,\% confidence level yielding only rough estimates of the parameters $a
=$ 1.26 (2.91), $p^\prime =$ 1.39 (.94) and $b =$ 1.14 (9.59).  Fits of similar
quality were found on the other side of the peak.  Simple logarithmic
plots are shown in Figs.~18 and 19 demonstrating consistency of the data
with a weak logarithmic divergence.  Clearly this ``brute force''
approach is not nearly as quantitative or decisive as the finite size
scaling study of the peak heights, but it is certainly compatible with
that data.  It is interesting that power-law fits, $C_h(\gamma, 16) =
a\mid\Delta\gamma\mid^{-\alpha} +$ b are not stable--the fitting
procedure always finds it can reduce the chi-squared of a fit by reducing
$\alpha$ while ``$a$'' grows positively and ``$b$'' grows negatively,
thus approximating a logarithm.

\section{Monopole Percolation}

{}~~~~~~We made a
detailed study of monopole percolation in the vicinity of the
Higgs-Coulomb transition studied quantitatively above. We used finite
size scaling methods since they have been so succesful in similar
studies done elsewhere. In particular,
lattices ranging from $6^4$ through $24^4$
were simulated at the $\gamma_{c}(L)$ values determined from the
specific heat peaks discussed above. The monopole order parameter $M$
and its associated susceptibility $\chi$ were then calculated over a
range of $\beta$ values. The results of these simulations on
$6^4$, $12^4$, and $18^4$ lattices are shown in Fig.~20, 21, and 22,
respectively. Typically, only 50,000-100,000 sweeps of the algorithm
were needed to obtain this data with their
relatively small error bars. We see
from the figures that a very clear percolation transition appears at
$\beta=.2325$. As was also found in our cruder simulations which
mapped out the phase diagram, the monopole percolation transition is
{\it not} coincident with the Higgs-Coulomb transition.

We can obtain several critical indices  of the percolation transition
by using scaling arguments.  For example, the peak of the monopole
susceptibility peak should depend on $L$ as,

$$
\chi_{max}\sim L^{\gamma/\nu} \eqno(11) $$

\noindent where $\gamma$ and $\nu$ are the susceptibility and
correlation length exponents. We test this scaling law in Fig.~23
where we see that the power law works very well with
$\gamma/\nu=2.25(1)$. This result is in excellent agreement with the
scaling law of ordinary four dimensional percolation, $9/4$$^{19.}$.
Since the bulk specific heats are not critical at this point, it is not
surprising that the monopole percolation indices would be uneffected
by the Higgs field. We attempted other measurements of
critical indices at the
percolation point, but they proved to be less quantitative.

\section{Four Dimensional Planar Model}

{}~~~~~~To check our results for the full theory, we confirmed  that our
techniques were able to reproduce known results.  For example, when
$\beta\rightarrow\infty$ Eq.~(1) reduces to the four dimensional
planar spin model which should have an order-disorder transition as a
function of $\gamma$ that is  described by mean field theory
$(\alpha=0, \nu=.5, etc.)$ with logarithmic corrections calculable in
perturbation theory$^{20.}$ -- for example, the index $p^\prime$ in Eq.(10)
is predicted to be $1/5$ for the O(2) model.
We measured the specific heat at the transition
for $L=6$, 8, 10, 12 and 14, and found peak values 20.47(3), 22.80(5),
24.35(9), 25.38(9)  and 26.24(9), respectively.  One million sweeps of
our code, tailored for $\beta=\infty$, were  run in each case.  We
note that for $L > 6$, the specific heat peaks grow with $L$ at a rate
for which is almost identical to the specific heat peaks
in the full theory. This numerical result is consistent with the
perspective developed above -- introducing
the gauge coupling in the model does not change the theory qualitatively.
The specific heat data are shown
in Fig.~24.
We also checked that the
correlation length exponent $\nu$ for this limit of SQED is
compatible with mean field theory. The peaks in the specific heat
occurred at $\gamma=.1556(1)$ at $L=6$, $.1541(1)$ at $8$, $.1532(1)$
at $10$, $.1526(1)$ at $12$, and $.1523(1)$ at $14$. As shown in
Fig.~25, these measurements are perfectly compatible with the scaling
law $\gamma_{c}(L)=aL^{-1/\nu}+b$ and the mean field value $\nu$=$1/2$.
As was the case in SQED, the data is not quite accurate enough
to search for the logarithms of Eq.(7).
And finally, in Fig.~26 we show the Kurtosis plot for the planar model
and see that it is compatible with the value $2/3$ for large $L$.
Certainly much more exacting studies of this model could be made (cluster
algorithms) and much larger lattices could be simulated,
but we are testing here just the simulation and analysis
technology available to the gauge model. The success of this test  study
gives us confidence that our SQED conclusions are reliable and the
logarithmic violations of mean field theory in SQED are real.

\section{Concluding Remarks}

{}~~~~~~One of the
motivations for this study was the recent finding that the chiral
symmetry breaking transition in non-compact lattice electrodynamics
with dynamical fermions is consistent with an ultra-violet stable fixed
point.$^{3.}$  Powerlaw critical behavior has been found with
non-trivial critical indices satisfying hyperscaling.  The present
negative result for scalar electrodynamics suggests that the chiral
nature of the transition for fermionic electrodynamics is an essential
ingredient for its scaling behavior.  It remains to be seen, however,
if the chiral transition found in fermion noncompact lattice QED
produces an interesting continuum field theory.

In conclusion, our numerical results support the notion that scalar
electrodynamics is a logarithmically trivial theory.  We suspect that
this result could be made even firmer by additional simulation studies
which use more sophisticated techniques such as renormalization group
transformations$^{5.}$ or partition function methods.$^{18.}$ We
are also hopeful that the data in Table 2. can be better organized
and exploited than we did here, and the presence of logarithmic
scaling violations can be extracted more quantitatively from this
finite size study.
Since we did not wish to bias our study toward logarithmic triviality,
we did not pursue special methods which require additional theoretical
input in order to be quantitative. However, it now seems appropriate
to execute a study of this type.$^{18.}$
Certainly our concentration on a
line of fixed electric charge in the entire phase diagram should be
relaxed.  Hopefully, accelerated Monte Carlo algorithms could be
developed~for scalar electrodynamics so that larger systems could be
simulated with better control.

We have presented calculations and fits to simulation data
on a wide range of lattice sizes and couplings, but since
the logarithms of interest are so slowly varying, we are skeptical
that our determinations of the exact powers of the various logarithms
are very quantitative. In particular, we know from
perturbative studies of $\lambda\phi^{4}$ that logarithmically
divergent specific heat peaks are accompanied by additive corrections
that fall away very slowly, as a small negative power of the logarithm.
It would take a wider range of couplings and
lattices to accommodate such nonleading terms meaningfully
into our fits, and once that could
be done, we suspect that the powers of the leading logarithmic
singularities discussed here could change quite significantly. Greater
analytic insight into SQED, or much more penetrating data analysis
methods appear necessary to quantitatively determine the powers of the
logarithms of interest with confidence. Nonetheless, we feel that
the {\it primary} goal of this research project was achieved --
SQED is compatible with logarithmic triviality and powerlaw critical
behavior indicative of a nontrivial ultra-violet stable fixed point
has no support.

\section*{Acknowledgement}

{}~~~~~~The simulation done here used the CRAY
C90's at PSC and NERSC.  We thank these centers for friendly user
access.  Several thousand cpu hours were needed to accumulate the
statistics listed in Table 1. The authors thank D. K. Sinclair
for help in the early stages of this work and acknowledge his assistance
in solving the random number problems discussed in the text.
J.B.K. is supported in part by the
National Science Foundation grant NSF PHY92-00148. S.K. is
supprted by DOE contract W-31-109-ENG-38. M.B. and H.F. acknowledge
the support of CESCA, CIEMAT and CICYT (project AEN \#93-0474).  H.F.
acknowledges support from CEE.

\section*{References} \begin{enumerate}

\item M. Baig, H. Fort, J.Kogut, S. Kim, and D. K. Sinclair, Phys. Rev.
{\bf D45}, R2385 (1993).

\item L. D. Landau and I. Ya. Pomeranchuk, Dokl. Akad. Nauk. {\bf
102}, 489 (1955).

\item S. Hands, A. Kocic, J. Kogut, R. Renken, D. K. Sinclair, and K.
C. Wang, Nucl. Phys. {\bf B413}, 503 (1994).

\item M. Baig, E. Dagotto, J. Kogut and A. Moreo, Phys. Lett. {\bf
B242}, 444 (1990).

\item D. Callaway and R. Petronzio, Nucl. Phys. {\bf277B}, 50 (1980).

\item J. L. Alonso et al., Zaragoza preprint, Sep 25, 1992.

\item C. N. Leung, S. T. Love, and W. A. Bardeen, Nucl. Phys. {\bf
B323}, 493 (1989).

\item S. Hands, and R. Wensley, Phys. Rev. Lett. {\bf 63}, 2169 (1989).

\item T. Banks, R. Meyerson, and J. Kogut, Nucl. Phys. {\bf B129}, 493
(1977). T. A. DeGrand and D. Toussaint, Phys. Rev. {\bf D22}, 2478
(1980).

\item M. N. Barber, in \underline{Phase Transitions and Critical
Phenomena}, Vol. VIII, eds. C. Domb and J. Lebowitz (Academic Press,
New York:  1983).

\item E. Brezin, J. Physique {\bf 43}, 15 (1982).

\item J. Rudnick, H. Guo and D. Jasnow, J. Stat. Phys. {\bf 41} 353
(1985).

\item M. Falcioni, E. Marinari, M. L. Paciello, G. Parisi and B.
Taglienti, Phys. Lett. {\bf 108B} 331 (1982).

\item A. M. Ferrenberg and R. H. Swendsen, Phys. Rev. Lett. {\bf 61},
2635 (1988).

\item M. Creutz, Phys. Rev. {\bf D36}, 515 (1987).

\item A. M. Ferrenberg, D. Landau and Y. J. Wong, Phys. Rev. Lett. {\bf
69}, 3382 (1992).

\item K. Binder, M. Challa and D. Landau, Phys. Rev. {\bf B34}, 1841
(1986).

\item R. Kenna and C. B. Lang, Phys. Lett. {\bf 264B}, 396 (1991).

\item S. Hands, A. Kocic and J. Kogut, Phys. Lett. {\bf 289B}, 400 (1992).

\item E. Brezin, J. C. Le Guillou and J. Zinn-Justin, Phys. Rev. {\bf B8},
2418 (1973).

\end{enumerate}

\section*{Figure Captions} \begin{enumerate}

\item The phase diagram of non-compact scalar electrodynamics.

\item A $\beta=0.1$ scan of the internal energies.

\item Same as Fig. 2, except at $\beta=0.2$.

\item Contour plot of $E_{\gamma}$.

\item Same as Fig. 4, except $E_{h}$.

\item Phase diagram on $6^4$ lattice showing line of specific heat
peaks and line of monopole susceptibility peaks (dashed).

\item Specific heat histogram on a $6^4$ lattice.

\item Same as Fig. 8, except on $12^4$ lattice.

\item Same as Fig. 9, except on $18^4$ lattice.

\item The Kurtosis $K_{\gamma}(L)$ vs. $10^{6}/L^{4}$.

\item Same as Fig. 11, except for $K_{h}$.

\item The specific heat peaks $C_{\gamma}^{max}(L)$ vs. $L$.  The solid
line is the logarithmic fit discussed in the text.

\item Same as Fig. 13., except for $C_{h}^{max}(L)$.

\item The critical coupling $\gamma_{c}(L)$ vs. $L^{-2}$.

\item Specific heat peaks vs. $\gamma$ on $12^4, 16^4, 20^4$ lattices.

\item Universal specific heat plot for powerlaw scaling.

\item Same as Fig.~17 except with scale breaking logarithm.

\item Coupling constant dependence of specific heat on $16^4$ lattice,
$\gamma < \gamma_c$ (16).

\item Same as Fig.~19, except $\gamma < \gamma_c$ (16).

\item The monopole susceptibility and order parameter plotted vs.
$\beta$ on a $6^4$ lattice.

\item Same as Fig. 16., except on a $12^4$ lattice.

\item Same as Fig. 16., except on a $18^4$ lattice.

\item Plot of ln$\chi_{max}$ vs. ln$L$.

\item Plot of ln$C_{max}$ vs. ln$L$, in the planar model.

\item Plot of $\gamma_{c}(L)$ vs. $1./L^{2}$, in the planar model.

\item Kurtosis plot for the planar model.

\end{enumerate}

\newpage

\begin{table}[a]    \caption{Finite Size Study of Scalar
Electrodynamics}   \vspace{.5 in}
\begin{tabular}{lllllll} L & $\gamma_c(L)$ & $C_{h}^{max}(L)$ &
$K_{h}(L)$ & $C_{\gamma}^{max}(L)$ & $K_{\gamma}(L)$ & Sweeps(millions)
\\
 6 & .23815(1)      & 13.81(2)          & .657668(9)  & 7.965(9)
& .665784(2) & 40\\
 8 & .23375(3)      & 15.83(2)          & .662954(5)  & 8.083(3)
& .666374(1) & 60\\ 10 & ..23173(1)      & 17.23(4)          &
.664892(4)  & 8.285(6)       & .666544(1) & 60\\ 12 & .23070(1)      &
18.43(7)          & .665713(4)  & 8.457(9)       & .666606(1) & 30\\
14 & .23004(1)      & 19.38(9)          & .666110(3)  & 8.594(15)
& .666633(1) & 20\\ 16 & .22962(1)      & 20.25(13)         &
.666319(2)  & 8.747(17)      & .666647(1) & 12\\ 18 & .22933(1)      &
20.85(15)         & .666441(2)  & 8.863(26)      & .666654(1) & 12\\
20 & .22912(1)      & 21.76(20)         & .666510(2)  & 8.956(20)
& .666658(1) & 10 \\  \end{tabular} \end{table}

\begin{table}[b] \caption{Specific Heat Data on $12^4, 16^4, 20^4$
Lattices}
\vspace{.5 in}
\begin{center}
\begin{tabular}{llll}
$\gamma$ $\qquad$&  $C_h$(12) $\qquad$&  $C_h$(16) $\qquad$&  $C_h$(20)\\
.2278     $\qquad$&   ---      $\qquad$&   ---     $\qquad$&   5.58(8)\\
.2280     $\qquad$&  ---       $\qquad$&  ---      $\qquad$&   5.99(6)\\
.2282     $\qquad$&  ---       $\qquad$&   ---     $\qquad$&   6.88(7)\\
.2284     $\qquad$&  ---       $\qquad$&   7.68(7) $\qquad$&   8.32(11)\\
.2286     $\qquad$&  ---       $\qquad$&   9.01(8) $\qquad$&  11.25(19)\\
.2288     $\qquad$&   ---      $\qquad$&  10.96(9) $\qquad$&  16.31(24)\\
.2290     $\qquad$&   ---      $\qquad$&  13.53(13)$\qquad$&  20.54(29)\\
.2292     $\qquad$&   ---      $\qquad$&  16.42(16)$\qquad$&  21.25(40)\\
.2294     $\qquad$&   11.56(5) $\qquad$&  19.07(16)$\qquad$&  18.51(29)\\
.2296     $\qquad$&   12.95(5) $\qquad$&  20.07(46)$\qquad$&  16.36(14)\\
.2298     $\qquad$&   14.41(5) $\qquad$&  19.49(20)$\qquad$&  15.16(13)\\
.2300     $\qquad$&   15.82(5) $\qquad$&  17.90(15)$\qquad$&  14.30(22)\\
.2302     $\qquad$&   17.03(5) $\qquad$&  16.48(9) $\qquad$&   ---\\
.2304     $\qquad$&   17.88(6) $\qquad$&  15.29(10)$\qquad$&   ---\\
.2306     $\qquad$&   18.27(8) $\qquad$&  14.57(8) $\qquad$&   ---\\
.2308     $\qquad$&   18.18(14)$\qquad$&  13.99(8) $\qquad$&   ---\\
.2310     $\qquad$&   17.99(16)$\qquad$&  13.79(12)$\qquad$&   ---\\
.2312     $\qquad$&   17.31(14)$\qquad$&   ---     $\qquad$&   ---\\
.2314     $\qquad$&   16.52(11)$\qquad$&   ---     $\qquad$&   ---\\
.2316     $\qquad$&   15.72(9) $\qquad$&   ---     $\qquad$&   ---\\
.2318     $\qquad$&   14.91(8) $\qquad$&   ---     $\qquad$&   ---\\
.2320     $\qquad$&   14.31(7) $\qquad$&   ---     $\qquad$&   ---\\
\end{tabular} \end{center}  \end{table}

\end{document}